\documentclass{article}

\usepackage[final]{neurips_2026}
 \usepackage{textcomp}

\usepackage{comment}
\usepackage{graphicx}
\usepackage{url}

\title{IRONSmith: A Visual Dataflow Design Environment for AMD Ryzen AI NPUs}

\author{
  Brock Sorenson, Samer Ali, Curt Bansil, Aman Arora \\
  Arizona State University \\
  \texttt{\{btsorens, swali6, cbansil, aman.kbm\}@asu.edu}
  \vspace{-10mm}
}

\begin{document}

% make the title area
\maketitle

% As a general rule, do not put math, special symbols or citations
% in the abstract
\begin{abstract}
\vspace{-3mm}
Machine learning inference increasingly relies on specialized hardware accelerators for throughput and power efficiency. Neural Processing Units (NPUs), such as the AMD Ryzen AI NPU, offer significant ML advantages over CPUs and GPUs, but programming them requires expertise in specialized frameworks. 
%The AMD Ryzen AI NPU uses IRON Python and mlir-aie, a tile-level spatial programming model requiring specification of inter-tile dataflow buffers (ObjectFIFOs), runtime data transfers, and compute tile kernel operations. Even experienced ML developers must invest substantial time in hardware documentation and programming guides before mapping a simple operation to the hardware. 
We present IRONSmith, the first visual dataflow design environment for programming AMD Ryzen AI NPUs. IRONSmith provides an interactive canvas displaying the AI Engine tile grid as visually connected blocks, allowing users to design ML dataflow applications by connecting tiles with wires representing FIFOs, split/join patterns, broadcast connections, and DDR transfers without writing any code. Compute kernels are assigned from a pre-built library, and worker functions are configured through property panels.
%Structural design verification checks designs for disconnected dataflow, DMA channel limit violations, and kernel argument mismatches before code generation, providing actionable feedback that would otherwise only surface as cryptic compiler or runtime failures. 
IRONSmith's backend pipeline automatically translates the visual design into executable IRON Python, handling structural completion, import resolution, and dependency management automatically. Generated code executes directly on the AMD Ryzen AI NPU. We demonstrate IRONSmith across ML designs of increasing complexity, from a single-tile vector passthrough to multi-tile matrix operations to a complete Multi-Layer Perceptron, all designed visually and successfully executed on the AMD Ryzen AI NPU. IRONSmith serves educators, students, ML researchers, and engineers by bridging the gap between ML knowledge and NPU programming expertise, widening access to hardware that is rapidly becoming standard across consumer and enterprise devices. 
%By bringing visual design principles to NPU programming, IRONSmith lowers the barrier to entry in much the same way that graphical development environments broadened access to software development.
\vspace{-3mm}
\end{abstract}

\section{Introduction} \label{sec:intro}
\vspace{-3mm}

Neural Processing Units (NPUs) have emerged as a critical hardware platform for efficient ML inference at the edge, offering significant throughput-per-watt advantages over CPUs and GPUs. The AMD Ryzen AI NPU is rapidly becoming standard across consumer laptops - yet programming it requires deep expertise in IRON Python and mlir-aie, a tile-level spatial programming model requiring manual specification of inter-tile FIFO buffers, runtime data sequences, and compute tile worker functions. 
%No visual design tool has existed for NPU programming; AMD's Riallto~\cite{riallto}, the closest prior effort, was discontinued and never offered a visual canvas or code generation.
We present IRONSmith, the first visual design environment for AMD Ryzen AI NPU programming. IRONSmith provides an interactive canvas where users wire tiles together with dataflow connections and configure each component through properties panels - all without writing code. Structural verification catches design errors before code generation, and the IRONSmith backend automatically produces complete, executable IRON Python. IRONSmith serves as both a learning tool for students in NPU architecture courses and an onboarding accelerator for ML developers beginning work with the Ryzen AI NPU.
\vspace{-3mm}

\section{Related Work} \label{sec:related_work}
\vspace{-3mm}

The mlir-aie project~\cite{mliraie} provides the open-source compiler infrastructure and the AMD IRON tutorial~\cite{iron2025}, which teach the programming model rather than lowering its barrier. AMD's Riallto~\cite{riallto} was the closest prior accessibility effort, offering Jupyter notebook tutorials for the spatial programming model, but was discontinued and never provided a visual canvas, design verification, or automated code generation. IRONSmith directly addresses the gap Riallto left.

Graphical programming has well-established precedent for lowering hardware entry barriers: Balid and Abdulwahed~\cite{balid2013} showed LabVIEW-based dataflow programming reduced FPGA development lifecycles by up to $5\times$, while Kuon et al.~\cite{kuon2014} and Winzker and Schwandt~\cite{winzker2019} demonstrated measurable student learning improvements with visual hardware tools. Commercial tools such as Matlab Simulink~\cite{simulink}, VisualApplets~\cite{visualapplets}, and AMD Vivado Block Design~\cite{vivado} apply this paradigm to FPGAs. For GPUs, NVIDIA Nsight~\cite{nsight} provides execution visualization but no design entry canvas. AMD Vitis~\cite{vitis} visualizes AI Engine mapping for Versal devices but requires C++ kernel authoring. No existing tool provides a visual design canvas for AMD Ryzen AI NPU programming. Ragan-Kelley et al.'s Halide~\cite{halide2013} established the principle of separating algorithm description from lowering and scheduling - a principle IRONSmith extends to the NPU domain, with the IRONSmith backend pipeline handling all scope, ordering, and dependency management automatically.

% INSERT COMPARISON TABLE HERE
\vspace{-3mm}

\section{Background} \label{sec:background}
\vspace{-3mm}

The AMD Ryzen AI NPU is built on the XDNA architecture: a spatial array of Shim tiles (DDR interface), Memory tiles (on-chip buffer storage), and Compute tiles (VLIW vector processors), connected exclusively through statically configured ObjectFifo buffers whose depths, types, and DMA channel counts must be resolved at design time. IRON Python~\cite{iron2025} programs this hardware by declaring ObjectFifos between tiles, defining runtime sequences that fill and drain DDR tensors through shim DMA engines, and implementing Compute tile Worker functions whose acquire/kernel-call/release loops must exactly match the declared ObjectFifo element types - meaning even a moderately complex design requires manually coordinating dozens of interdependent FIFOs, split/join patterns, and worker configurations.
%across hundreds of lines of code.

The fundamental challenge is that IRON Python expresses an inherently spatial programming model - a graph of tiles, FIFOs, and data movements - as a sequential text file, requiring the programmer to mentally track tile connectivity, hardware requirements, and dataflow across hundreds of lines of code. Structural errors such as disconnected FIFOs, DMA channel oversubscription, or kernel argument mismatches produce compiler errors that are difficult to diagnose without a visual representation of the design. IRONSmith eliminates this mismatch by making the spatial programming model directly visual, and catches the most common structural errors before code generation.

\vspace{-1mm}

\begin{figure}[htbp]
\centering
\includegraphics[width=\columnwidth]{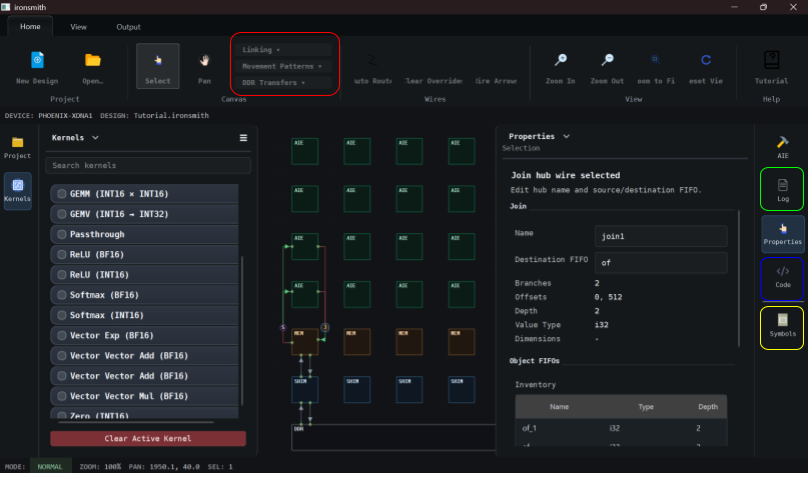}
\caption{Annotated IRONSmith GUI overview showing the AI Engine tile grid canvas (center), FIFO wire connected tiles, kernel library (left), properties panel (right), data movement patterns (red), generated code panel (blue), symbol table (yellow), and output log (green).}
\label{fig:gui}
\end{figure}
\vspace{-2mm}

\begin{figure}[htbp]
\centering
\includegraphics[width=\columnwidth]{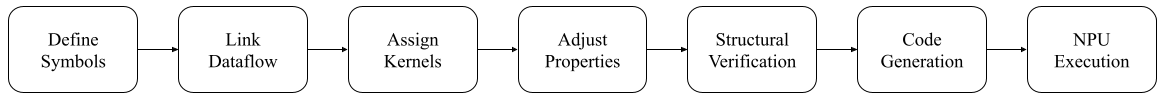}
\caption{IRONSmith design workflow}
\label{fig:workflow}
\vspace{-4mm}
\end{figure}

\section{IRONSmith} \label{sec:proposal}
\vspace{-3mm}

IRONSmith addresses the NPU programming barrier by replacing the text-based IRON Python workflow with a visual, interactive canvas (Figure \ref{fig:gui}). Users design NPU dataflow applications visually on a canvas and then generate executable IRON Python without writing a single line of code.  
IRONSmith is built on a microkernel plugin architecture in which a lightweight Core System manages the application lifecycle while modular plugins - Canvas, Project Explorer, Code Editor, etc. - contribute functionality through a central ExtensionSystem service registry, allowing new visual features or backend targets to be added without modifying the core.
IRONSmith is architected with scalability in mind - there is a frontend (the visual user-facing canvas) and a backend (the compiler that lowers the visual design into code).

% INSERT ARCHITECTURE DIAGRAM HERE

Building a design in IRONSmith follows a fixed seven-step workflow (Figure~\ref{fig:workflow}): users \textit{define symbols} in the Symbol Table, \textit{link dataflow} by wiring runtime, FIFOs, and other patterns between the canvas tiles, \textit{assign kernels} to each compute tile from the Kernel Selection Library, and \textit{adjust properties} - FIFO parameters, runtime sequences, worker functions, etc. - through the properties panel. Once the design is complete, users trigger \textit{structural verification}, then \textit{code generation}, producing executable IRON Python ready for \textit{NPU execution} on the AMD Ryzen AI NPU.

\vspace{-2mm}
\subsection{Canvas and Visual Design Layer}
\vspace{-3mm}

The canvas presents the AMD AI Engine tile grid as an interactive layout where users connect tiles with typed wires representing the full set of dataflow patterns: ObjectFifo (tile-to-tile), Split/Join (sub-FIFO memory tile access), Broadcast (one-to-many), and Forward (pass-through). A context-sensitive Properties Panel exposes all underlying parameters - FIFO depth, element type, dimensions, DDR sizes, Tensor Access Patterns, Kernel assignments, and Worker acquire/call/release sequences - through structured form fields. A Symbol Table registers named constants and type abstractions for consistent reuse across the design. A searchable kernel library sourced from mlir-aie~\cite{mliraie} lets users assign pre-built ML kernels (GEMM, ReLU, CONV2D, Softmax, etc.) per compute tile. Custom user-defined kernels can also be assigned to tiles through the GUI.
Designs are persisted automatically to a \texttt{document.json} bundle on every change.

\vspace{-2mm}
\subsection{Design Verification and Code Generation}
\vspace{-3mm}

Before generating code, IRONSmith runs structural checks catching the most common IRON Python error classes: invalid runtime I/O configuration, disconnected dataflow, DMA channel limit violations, and kernel argument mismatches - each producing an actionable output log message rather than a cryptic compiler failure.

The IRONSmith backend takes the \texttt{document.json} (the auto-saved persistent canvas representation) and instantiates a High-Level Intermediate Representation (HLIR): an in-memory builder API that constructs typed component objects for each tile, FIFO, kernel assignment, and runtime sequence in the design. The HLIR is then lowered through a staged pipeline - to a \texttt{.xml} file representation, then a semantically directed graph, and finally a complete, executable IRON Python file - with each stage resolving interdependencies, inferring missing values, enforcing IRON coding conventions, and determining the correct import set automatically.

% INSERT DESIGN LOWERING PIPELINE FIGURE HERE
\vspace{-2mm}

\section{Demonstration} \label{sec:methodology}
\vspace{-3mm}

\subsection{Benchmark Suite}
\vspace{-3mm}

We evaluate IRONSmith across three dimensions: correctness (proper execution of generated IRON Python), expressiveness (IRON dataflow patterns represented), and productivity (efficiency of design implementation). We developed seven benchmark designs of increasing complexity, each built visually in IRONSmith and executed successfully on the AMD Ryzen AI NPU (Table~\ref{tab:benchmarks}), progressing from a single-tile passthrough through element-wise vector operations to dense linear algebra with split/join and broadcast patterns.
All seven benchmarks generated correct IRON Python that executed on the hardware NPU without modification. The benchmark suite exercises every major ObjectFifo pattern supported by the IRON Python model.

\vspace{-2mm}
\begin{table}[h]
\centering
\caption{IRONSmith benchmark suite executed on AMD Ryzen AI NPU.}
\label{tab:benchmarks}
\small
\begin{tabular}{lccccl}
\hline
\textbf{Benchmark} & \textbf{Shim} & \textbf{Mem} & \textbf{Compute} & \textbf{FIFOs} & \textbf{Capabilities Demonstrated} \\
\hline
Passthrough   & 1 & 0 & 1  & 2  & ObjectFifo, linear fill/drain \\
Vector Exp    & 1 & 1 & 1  & 2  & Kernel selection, single worker \\
Vec-Vec Add   & 1 & 1 & 4  & 10 & Split/Join mem tile, multiple workers \\
Vec-Vec Mul   & 2 & 2 & 8  & 20 & Multi-shim, multi-column runtime \\
GEMV          & 2 & 2 & 8  & 24 & Broadcast, matrix/vector tiling \\
GEAM          & 4 & 4 & 16 & 60 & Dual matrix inputs, 16 tiles \\
GEMM          & 4 & 4 & 16 & 60 & Accum.\ in-place, custom core body, TAPs \\
\hline
MLP           & 4 & 4 & 16 & 42 & Tensor manipulation (\texttt{dims\_from/to\_stream}) \\
\hline
\end{tabular}
\vspace{-3mm}
\end{table}

\subsection{MLP Neural Network: Complex Use Case}
\vspace{-3mm}

To demonstrate IRONSmith's capability for complex, multi-layer ML designs, we implemented a complete Multi-Layer Perceptron (MLP) using the visual canvas. The MLP consists of 4 layers: 3 hidden layers each performing a 64$\times$64 matrix multiply followed by a ReLU activation, and one output layer performing a 64$\times$64 matrix multiply followed by a Softmax activation. All weight matrices are 64$\times$64 BFloat16 values (4096 elements per matrix), and the design maps onto 16 AMD AI Engine Compute tiles. The network is designed entirely within IRONSmith's canvas, with layers connected sequentially by dataflow wires and kernels assigned from the pre-built library.

% INSERT MLP CANVAS SCREENSHOT HERE
\vspace{-2mm}
\begin{figure}[htbp]
\centering
\includegraphics[width=\columnwidth]{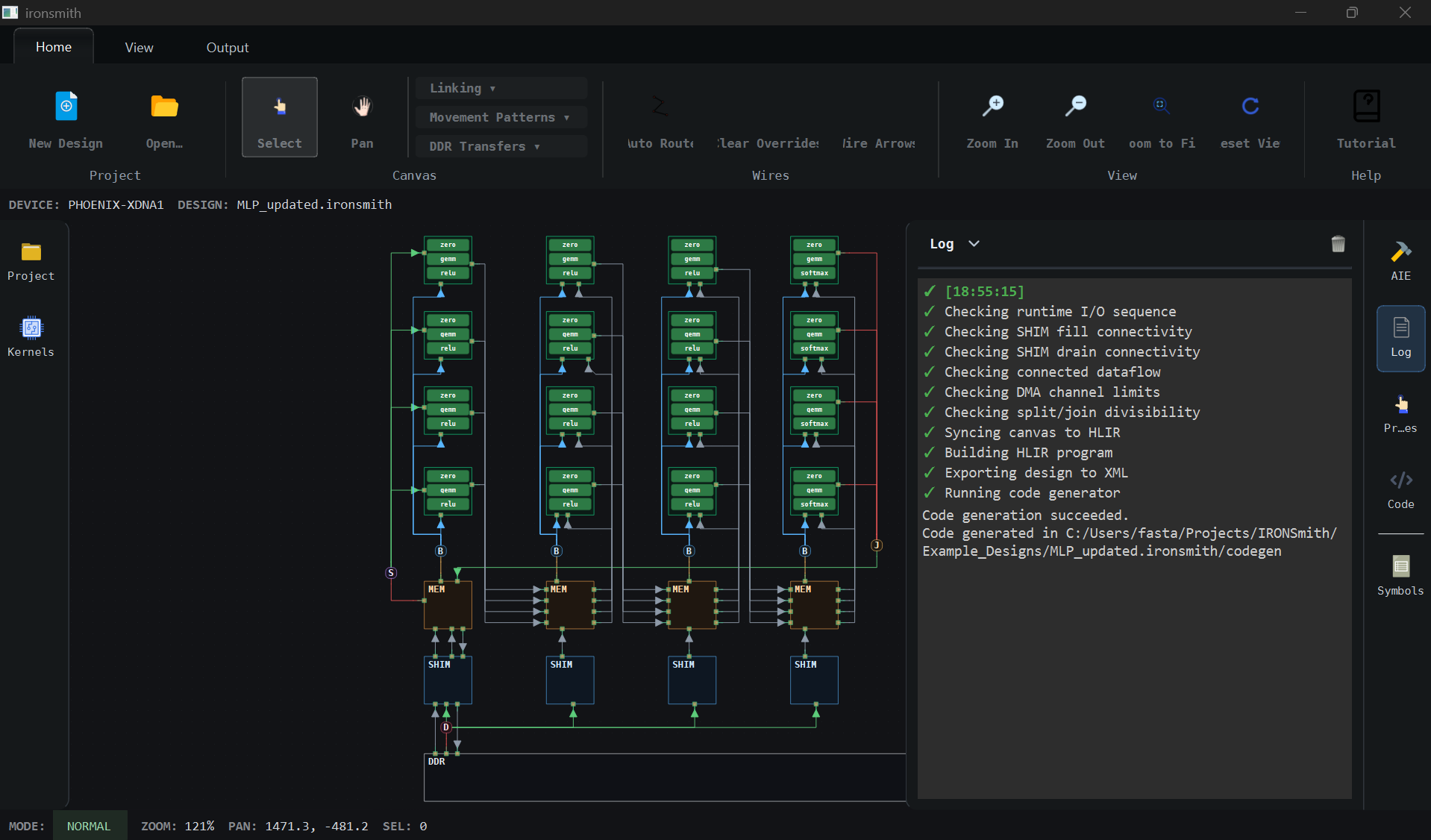}
\caption{IRONSmith canvas view of the MLP design, showing 3 ReLU activation layers and one output layer with 4 broadcast weight matrices.}
\label{fig:mlp}
\end{figure}

Figure~\ref{fig:mlp} shows the complete MLP canvas: 16 Compute tiles with 42 inter-tile FIFOs, 4 weight broadcast connections fanning out from DDR to each tile column, and inter-layer activation connections passing through memory tiles for tensor manipulation. The design passes all structural verification checks and generates correct IRON Python that executes on the AMD Ryzen AI NPU - a design that would require approximately 200 lines of manually coordinated IRON Python code, produced here without writing any.
\vspace{-2mm}

\section{Conclusion} \label{sec:conclusion}
\vspace{-3mm}

IRONSmith is the first visual dataflow design environment for AMD Ryzen AI NPU programming, providing a complete pipeline from graphical canvas design to automatically generated, directly executable IRON Python. We demonstrated IRONSmith across seven benchmark designs of increasing complexity - from a single-tile passthrough to GEMM - and a complete four-layer MLP, all successfully executed on the AMD Ryzen AI NPU. The benchmark suite covers every major IRON dataflow pattern, and the MLP demonstrates that IRONSmith scales to real ML applications without requiring users to write any code.

Future work includes expanding the pre-built kernel library, adding in-canvas performance profiling, and a quantitative user study measuring onboarding time against direct IRON Python programming. IRONSmith is open source (GPL v3.0) at \url{https://github.com/advent-lab/IRONSmith}, with user documentation and a getting-started guide available in the repository.

% trigger a \newpage just before the given reference
% number - used to balance the columns on the last page
% adjust value as needed - may need to be readjusted if
% the document is modified later
%\IEEEtriggeratref{8}
% The "triggered" command can be changed if desired:
%\IEEEtriggercmd{\enlargethispage{-5in}}

% references section

% can use a bibliography generated by BibTeX as a .bbl file
% BibTeX documentation can be easily obtained at:
% http://mirror.ctan.org/biblio/bibtex/contrib/doc/
% The IEEEtran BibTeX style support page is at:
% http://www.michaelshell.org/tex/ieeetran/bibtex/
%\bibliographystyle{IEEEtran}
% argument is your BibTeX string definitions and bibliography database(s)
%\bibliography{IEEEabrv,../bib/paper}
%
% <OR> manually copy in the resultant .bbl file
% set second argument of \begin to the number of references
% (used to reserve space for the reference number labels box)
%\begin{thebibliography}{1}
%
%\bibitem{IEEEhowto:kopka}
%H.~Kopka and P.~W. Daly, \emph{A Guide to \LaTeX}, 3rd~ed.\hskip 1em plus
%  0.5em minus 0.4em\relax Harlow, England: Addison-Wesley, 1999.
%
%\end{thebibliography}

\bibliographystyle{plainnat}
\bibliography{refs}

% that's all folks
\end{document}